\documentclass[preprint]{aastex}
\usepackage{amsmath, amsfonts,amssymb,natbib,graphicx,subfig,rotating,lscape,multirow,ulem,color}

\shorttitle{Simulated void galaxies in the CDM model}

\begin{document}

\title{Simulated void galaxies in the standard cold dark matter model}

\author{Kathryn Kreckel\altaffilmark{1}, M. Ryan Joung\altaffilmark{1}, and Renyue Cen\altaffilmark{2}}

\altaffiltext{1}{Department of Astronomy, Columbia University, Mail Code 5246, 550 West 120th Street, New York, NY 10027, USA}
\altaffiltext{2}{Princeton University Observatory, Princeton, NJ 08544, USA}


\begin{abstract}
We analyze a (120 $h^{-1}$ Mpc)$^3$ adaptive mesh refinement hydrodynamic simulation that contains 
a higher-resolution  $31 \times 31 \times 35 h^{-3}$ Mpc subvolume centered on a $\sim$30 Mpc diameter void.
Our detailed $\sim$1 kpc resolution allows us to identify 1300 galaxies within this void to a limiting halo mass of $\sim10^{10} M_\sun$.  Nearly 1000 galaxies are found to be in underdense regions, with 300 galaxies  residing in regions less than half the mean density of the simulation volume.  We construct mock observations of the stellar and gas properties of these systems, and reproduce the range of colors and luminosities observed in the SDSS for nearby ($z < 0.03$) galaxies.  We find no trends with density for the most luminous (M$_r < -18$) galaxies, however our dwarf void galaxies (M$_r > -16$), though they are less reliably resolved, typically appear bluer, with higher rates of star formation and specific star formation and lower mean stellar ages than galaxies in average density environments.  We find a   significant population of low luminosity (M$_r \sim -14$) dwarf galaxies that is preferentially located in low density regions and specifically in the void center.  This population may help to reduce, but not remove, the discrepancy between the predicted and observed number of void galaxies.
\end{abstract}

\section{Introduction} 
\label{sec:intro}

$\Lambda$CDM has proven fairly robust when compared with a wide range of existing observations \citep{Lacey1993, Cen1994, Zhang1995,  Springel2005, Komatsu2011}. Simulations can reproduce the galaxy clustering properties we observe in large redshift surveys, as well as the size of galaxy clusters. In addition, $\Lambda$CMD predicts hierarchical growth as the dominant mechanism in galaxy evolution \citep{Peebles1969, White1991}.

However, discrepancies with observation remain, particularly on individual galaxy scales.  Theory predicts a cuspy core dominate the shape of the dark matter density profile, while observations show that a wide range of halo sizes, from galaxy clusters to dwarf spheroidals, instead have cored profiles \citep{Tyson1998, Kleyna2003}.  Simulations have difficulty reproducing exponential, bulgeless disks, and observations within the Local Volume disagree with predictions of where we should find the largest disk galaxies \citep{Peebles2010, Kormendy2010}. The missing satellite problem describes from the overprediction of small dark matter halos clustering around the Milky Way Galaxy \citep{Klypin1999, Moore1999, Simon2007}. The void phenomenon refers to the overprediction of the number of low-mass halos existing within voids in cosmological simulations \citep{Peebles2001}. One key to solving these problems may come from the inclusion of sufficient physics to properly simulate the baryon content of galaxies, which does not dominate their mass but  contributes significantly to small scale dynamics and observables, such as stars or gas. 

Early cosmological simulations of voids were based on dark matter only, N-body simulations (e.g., \citealt{Ryden1984, White1987, Little1994, Vogeley1994, Mathis2002}).  These authors compared the dark matter distribution with the void statistics in large surveys such as the CfA redshift catalogs and the Las Campanas redshift survey, in terms of their sizes and abundances.  Using semi-analytic prescriptions for galaxy formation on outputs from N-body simulations, \cite{Patiri2006b} studied the colors and  specific star formation rates of void galaxies, finding that there are more blue galaxies in void but no systematic differences between void galaxies and the general galaxy population.
 \cite{Ceccarelli2006} used a similar method to study void dynamics and the effect of redshift distortions in void identification.
  The only cosmological hydrodynamic simulations run to date focusing on voids are \cite{Viel2008}, who analyzed the
void statistics at $z \sim 2$, and \cite{Hoeft2006}, who investigated the effect of the cosmological UV background on the formation of dwarf galaxies in voids.

We have selected a void region from within a full (120 $h^{-1}$ Mpc)$^3$ cosmological simulation, and we examine with moderate resolution the dark matter, gas and stars of a large sample of galaxies located within and around this large void. We employ an adaptive mesh refinement (AMR) code to reproduce the gas physics on a scale of 1 kpc. Observationally, void galaxies are distinguishable from galaxies in average or overdense environments as they are typically less luminous, and at fixed luminosity are bluer, with higher rates of star formation and specific star formation \citep{Rojas2004, Rojas2005}.  This is not strongly reflected in their gas content, though they are generally gas-rich, as their total mass in H \textsc{i} is fairly typical for their luminosities \citep{Szomoru1996, Kreckel2011}.

We compare the integrated properties of void galaxies from our simulation, described in Section \ref{sec:simulation}, both with observed void galaxies and galaxies in higher density regions within this simulation.  We present our results in Section \ref{sec:results} and our conclusions are summarized in Section \ref{sec:conclusion}.

\section{Simulation Initial Conditions and Physical  Processes}
\label{sec:simulation}

We perform cosmological simulations using the adaptive mesh refinement
Eularian hydro code, Enzo \citep{Bryan1999, Norman1999, Oshea2004}, with a
periodic box of size 120 $h^{-1}$ Mpc comoving on a side and cosmological
parameters taken from the WMAP5 $\Lambda$CDM results combined with measurements of Type Ia supernovae and Baryon Acoustic Oscillations in the galaxy distribution \citep{Hinshaw2009}
: ($\Omega_m, \Omega_\Lambda,\Omega_b,h,\sigma_8,n_s$) = (0.279, 0.721, 0.0462, 0.701, 0.817, 0.960).
We first run a low-resolution simulation with a uniform dark matter
particle mass of 7.6 $\times 10^{10} ~M_\sun$, 128$^3$ root grid cells,
and only 4 levels of refinement from an initial redshift of $z=99$ to 0.
Based on the $z=0$ output of this simulation, a large void region having a
diameter of $\sim$30 $h^{-1}$ Mpc was identified. (see also \cite{Cen2010}).  
Tracing the dark matter
particles back to the initial redshift of $z=99$ showed that the void
region expands with time in terms of comoving volume, as expected.  To
achieve high mass and spatial resolution, we then use the multimass
initialization technique and employ three nested volumes with successive
particle masses decreased by a factor of 8.  Hence, the innermost $31 \times 31 \times 35h^{-3}$Mpc comoving volume, 
sufficiently large to contain the identified
void volume at $z=0$, has a dark matter particle mass of 1.5 $\times
10^8~M_\sun$.
Within this innermost nested volume, hydrodynamic refinements were allowed
beginning with a root grid cell size of 937 $h^{-1}$ kpc and a maximum
refinement level of $l_{\rm max}$ = 10, resulting in a maximum resolution of
0.916 $h^{-1}$ kpc at $z=0$.

The simulation includes a metagalactic UV background
(\citealt{Haardt1996}), a diffuse form of photoelectric and
photoionization heating \citep{Abbott1982, Joung2006}, and
shielding of UV radiation by neutral hydrogen \citep{Cen2005}.   They
also include cooling due to molecular hydrogen \citep{Abel1997} and
metallicity-dependent radiative cooling \citep{Cen1995} extended down to
10 K \citep{Dalgarno1972}.  

Inside the innermost nested volume where
AMR is allowed, additional physical processes of star formation and
feedback are implemented.  Star particles are created in cells that
satisfy a set of criteria for star formation proposed by \cite{Cen1992}, which requires that the gas within that cell be contracting, cooling rapidly and gravitationally unstable. 
Under these conditions, a stellar particle of mass
$m_* = c_* m_{\rm gas} \Delta t / t_*$ is created to replace gas from that cell, and is tagged with its initial mass, creation time, and metallicity.
Here, $\Delta t$ is the time step, $t_*$ = max($t_{\rm dyn}, 3 \times 10^6 {\rm yr})$, $t_{\rm dyn} = \sqrt{3 \pi / (32 G \rho_{\rm tot})}$ is the dynamical time of the cell, $m_{\rm gas}$ is the baryonic gas mass in the cell, and $c_* = 0.03$ is the star formation efficiency.  
Star particles typically  have an initial mass of $\sim 5 \times 10^6 M_\sun$. 
Star formation and supernovae feedback are modeled following \cite{Cen2005} with a supernovae efficiency of $e_{SN} = 10^{-5}$. Star particle masses decay slightly as feedback energy and ejected metals are distributed into the 27 local gas cells, weighted by the specific volume of each cell, and centered at the star particle in question.  The temporal release of metal enriched gas and thermal energy at time $t$ has the following form: $f(t,t_i,t_{\rm dyn}) \equiv (1 / t_{\rm dyn})[(t-t_i)/t_{\rm dyn}] \exp[-(t-t_i)/t_{\rm dyn}]$, where $t_i$ is the formation time of a given star particle. 
The metal enrichment inside galaxies and in the intergalatic medium
(IGM) is followed self-consistently in a spatially resolved fashion \citep{Cen2005}.  Recently, \cite{Cen2010} used a simulation of the same system
with a higher spatial resolution (by a factor of 2) to study the nature of
damped Ly-$\alpha$ systems.  

\section{Mock Observations within the Simulation Volume}
\label{sec:mockobs}
We identify virialized objects in our
high-resolution simulations using the HOP algorithm \citep{Eisenstein1998} with the threshold parameter ($\delta_{\rm outer}$) of 125.  For
identifying galaxies, we used higher $\delta_{\rm outer}$ values of 10$^3$
and 10$^4$ to additionally find subhalos located within virialized
objects.
Galaxies are excluded that are outside of the central, high-resolution $\sim$ (30 $h^{-1}$ Mpc)$^3$ region, or that contain any coarse dark matter particles more massive than the highest particle resolution achieved. We also exclude any dark matter halos with fewer than 100 particles within the virialized region to ensure sufficient resolution of the physical balance between pressure and self-gravity within the simulated galaxies.  For this work, we have extracted the hydrodynamic data around every identified galaxy at a uniform resolution corresponding to level 8,
or a physical scale of 3.66 $h^{-1}$ kpc at $z=0$, as it is sufficient for our examination of the integrated properties in these systems and significantly speeds our analysis.

The light distribution is computed from the star particles using the Galaxy Isochrone Synthesis Spectral Evolution Library (GISSEL) stellar
synthesis code \citep{Bruzual2003}.  We calculate the luminosities of the simulated galaxies in the five Sloan Digital Sky Survey (SDSS, \citealt{Abazajian2009}) bands ($ugriz$).  
Internal extinction corrections were applied by considering the mass column density of metals, $\Sigma_Z$, along the line of sight within the virial radius for a randomly chosen viewing direction along the simulation volume axes. Starting with the observational relation determined in the Milky Way at solar metallicity \citep{B&M},
\begin{equation}
\label{eq:avobs}
A_V =  \frac{N(H_{\rm tot})}{1.9 \times 10^{21} ~{\rm cm}^{-1}} ~mag ,
\end{equation}
we allow for some dependence on metallicity, and scale for the fraction of refractory elements, $f_{Fe}$ \citep{Vladilo2004}. Thus we find that 
\begin{equation}
\label{eq:avsim}
A_V = \frac{\Sigma_Z ~f_{\rm Fe}}{F ~m_p ~4 \times 10^{19} ~{\rm cm}^{-2}} ~mag .
\end{equation}
 We choose a scaling factor, $F=1.5$, to match our simulated extinction with the observational relation (Equation \ref{eq:avobs}) for those simulated galaxies with solar metallicity.  
We also considered the observational dust to gas relation determined in the Large Magellanic Cloud, which has a much larger uncertainty, as compared to 1/3--1/2 solar metallicity simulated galaxies \citep{Koornneef1982}. We find a best fit with a factor of $F\sim3$, however a factor of $F = 1.5$ is still in agreement to within the errors, which are quite large, and we adopt this value for the remainder of this paper.
 Extinctions for each star particle were then scaled to the central wavelength for each of the five SDSS bands following \cite{Calzetti2000}.
 We define the integrated
stellar parameters (i.e. mass, luminosity, star formation rate) of each simulated galaxy to be the sum of the properties of the star particles located within 15\% of the virial radius of the galaxy at a given redshift.

The dark matter density field was determined by calculating the mean dark matter density for the entire simulation volume,  gridding all dark matter particles to a  1 $h^{-1}$ Mpc grid, then applying a 5 $h^{-1}$ Mpc three-dimensional Gaussian smoothing filter.  The smoothing length was chosen to agree with the galaxy correlation length observed in large redshift surveys \citep{Jing1998, Park2007}.  We also note that our choice of smoothing length assigns a reasonable dark matter density contrast of $\delta\rho/\rho < -0.5$ to all galaxies where the distance to the third nearest neighbor is greater than 7 $h^{-1}$ Mpc, which was the more restrictive void galaxy selection technique used by  \cite{Rojas2004}.  


\section{Results} 
\label{sec:results}
\begin{figure}
\centering
\includegraphics[angle=0,width=4in]{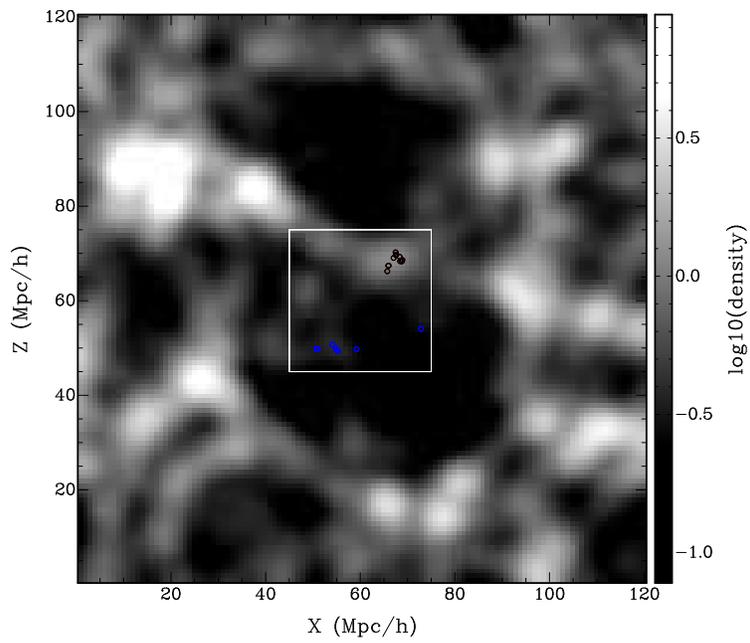}
\caption{Slice of thickness 1 $h^{-1}$ Mpc through the simulation volume smoothed dark matter density field.  The high resolution region (white box) contains a large fraction of the large central void, and the positions of NV (black circles) galaxies trace the higher density regions while the LVS (blue circles) galaxies are located well within the void. \label{fig:xyz}}
\end{figure}

\begin{figure}
\centering
\vspace{-1.5cm}
\includegraphics[width=3in]{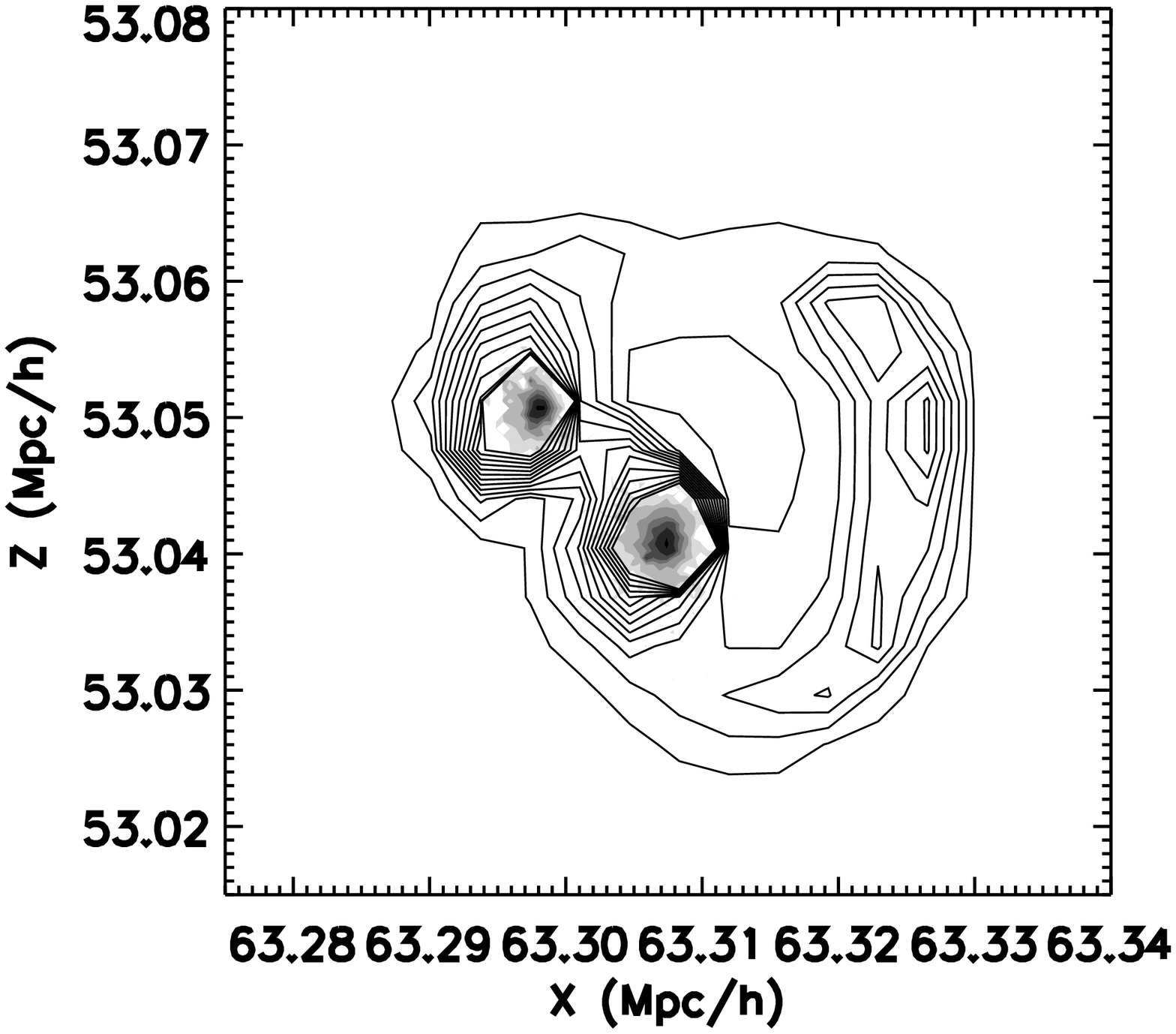}
\includegraphics[width=2.7in]{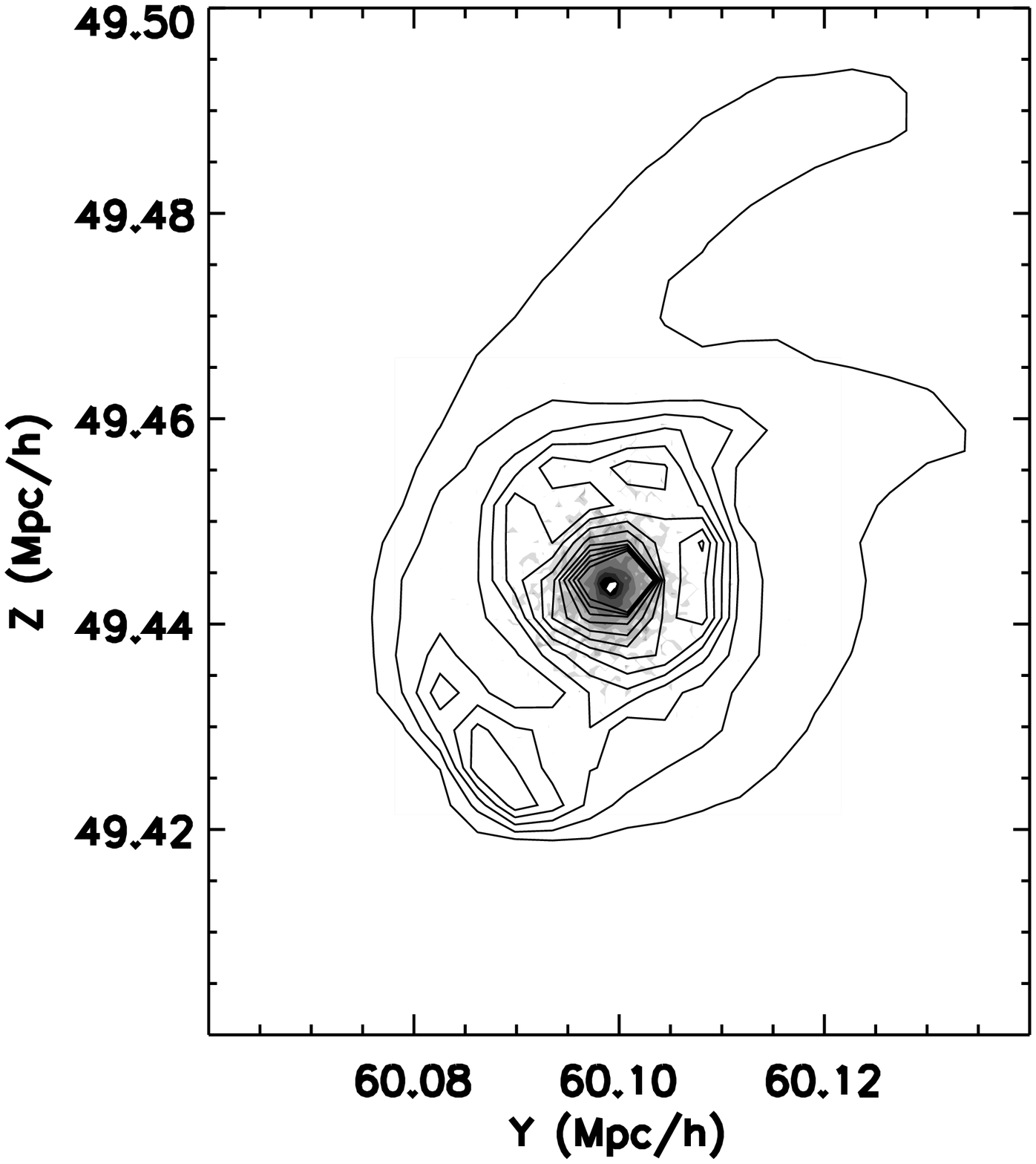}\\
\vspace{-3cm}
\includegraphics[width=3in]{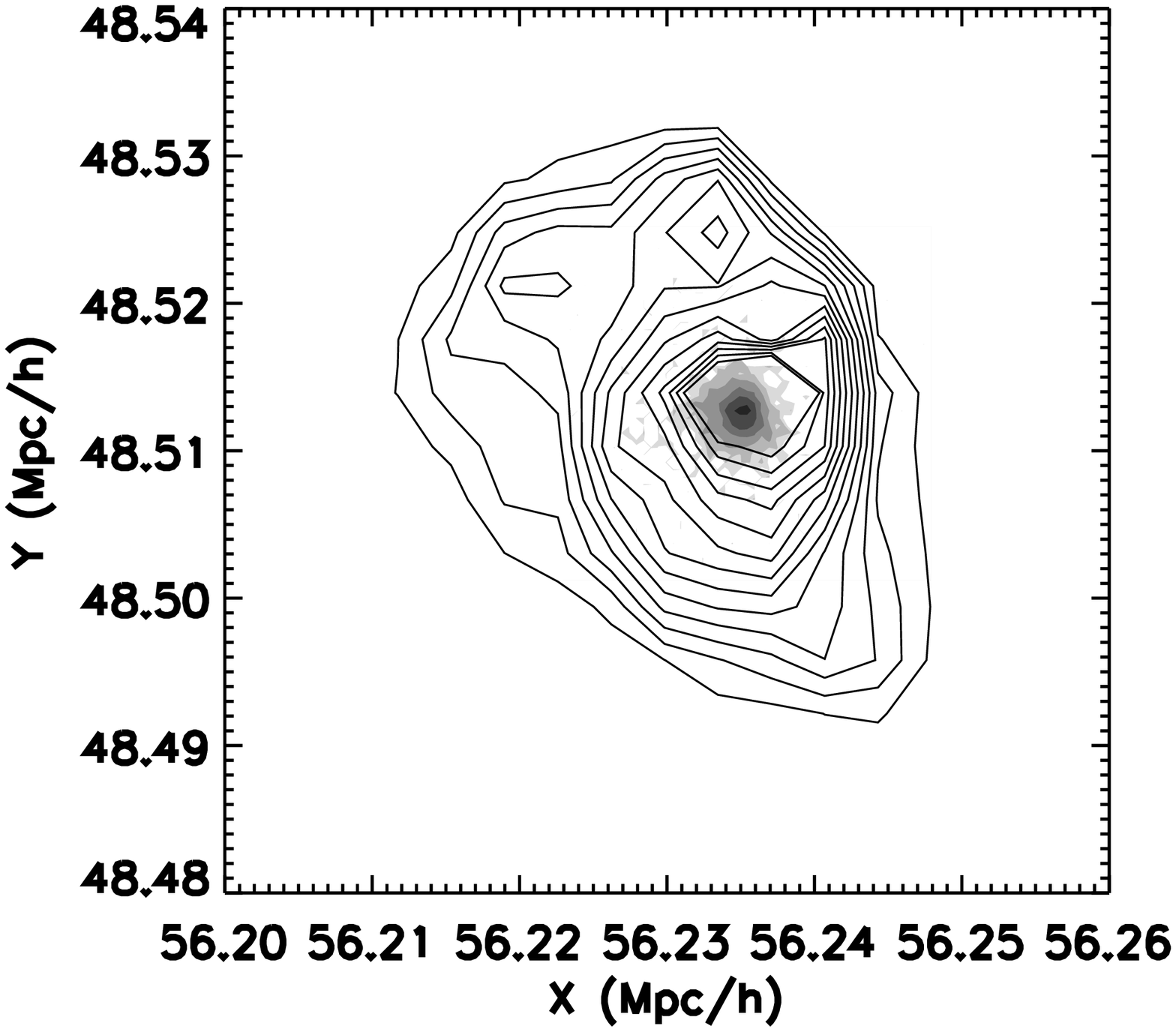}
\includegraphics[width=3in]{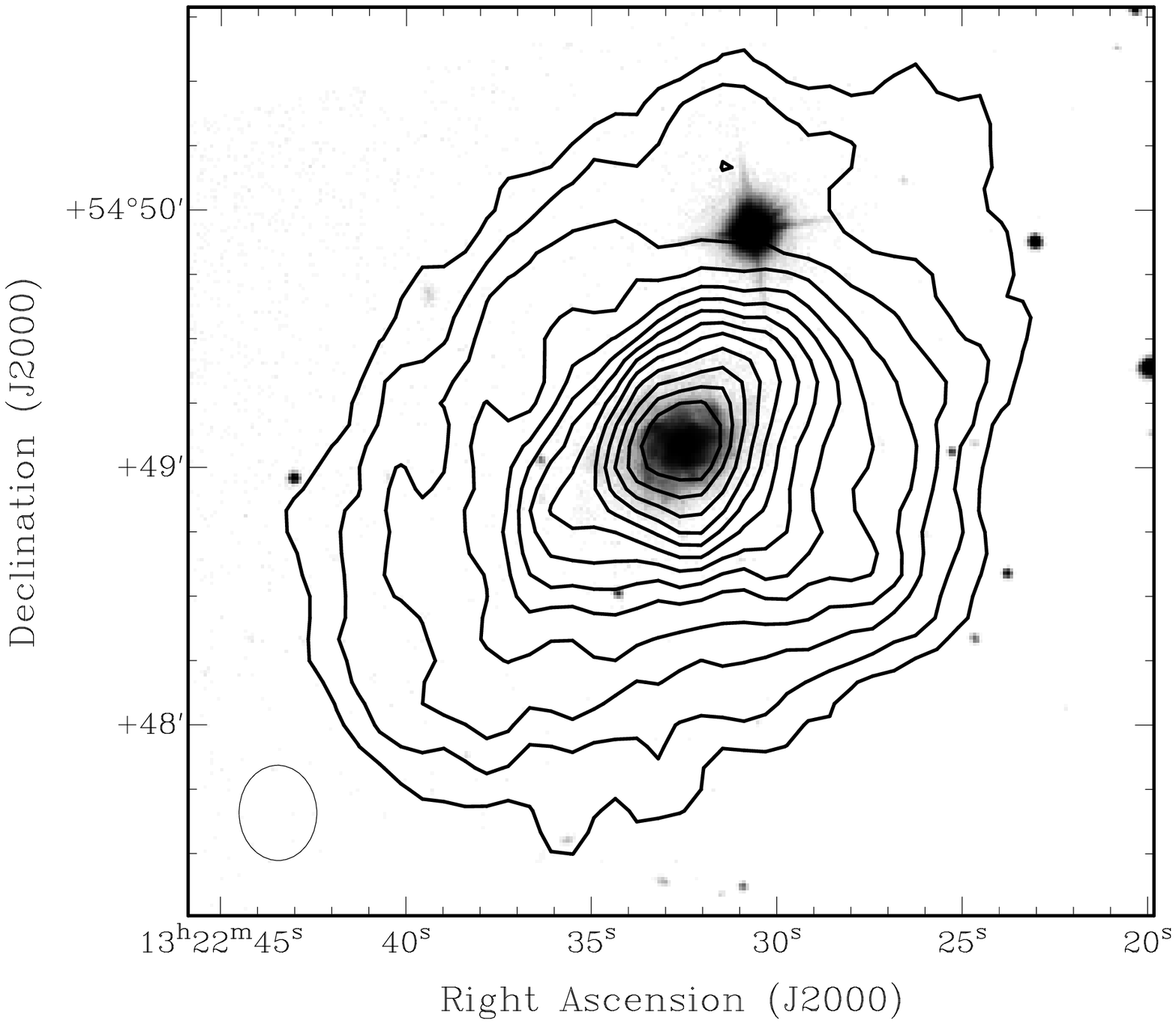}
\caption{Simulated observations of LVS galaxies, with sample observed void galaxy (bottom right) for reference. Greyscale indicates g-band emission, contours show H \textsc{i} column densities of $5 \times 10^{19}~{\rm cm}^{-2}$ plus increments of $10^{20} ~{\rm cm}^{-2}$ with a maximum of $1.25 \times 10^{21} ~{\rm cm}^{-2}$.  The observed void galaxy VGS\_32 (bottom right), described in \cite{Kreckel2011}, shows the H \textsc{i} beam shown in the lower left.  The simulated and observed galaxies have roughly the same luminosity and H \textsc{i} extent, and in general the agreement is good.  However, the simulated galaxies have significantly more low column density H \textsc{i}.
\label{fig:simobs}}
\end{figure}

Of a total of 1268 galaxies with high resolution hydrodynamic simulation, none are found at positions with a dark matter density contrast (see Section \ref{sec:mockobs}) of more than 1 as we include only filaments and walls bounding the void but no massive clusters.  950 galaxies are found at locations with $\delta\rho/\rho < 0$, from which we form a void sample (VS) of 648 galaxies with $-0.5 < \delta\rho/\rho < 0$, and 302 that are in regions with $\delta\rho/\rho < -0.5$ form a low-density void sample (LVS).  There are also 318 galaxies that make up a non-void sample (NV) at roughly average densities ($0 < \delta\rho/\rho < 1$) on the void boundaries (Table \ref{tab:cats}, Figure \ref{fig:xyz}).  
We make these divisions in part to facilitate direct comparison with observed void samples \citep{Rojas2004, Rojas2005}.
We can also define the distance, R,  from the void center at [64, 71, 45] $h^{-1}$ Mpc, as identified within the full 120 $h^{-1}$ Mpc box, and find that only two  galaxies resolved by the simulation are within 5.5 $h^{-1}$ Mpc and no NV galaxies are within 18 $h^{-1}$ Mpc.  We use this to additionally define a void center sample (VC) of 176 galaxies within 18 $h^{-1}$ Mpc. 
Simulated observations were made with a typical H \textsc{i} column density detection limit of $1 \times 10^{19} {\rm cm}^{-2}$, and resulting integrated H \textsc{i} column density contours overlaid on simulated observations of the stellar luminosity (Figure \ref{fig:simobs}).

We apply a Kolmogorov-Smirnov test to determine if the luminosity distribution of these samples could be drawn from the same parent population.  The probability that the LVS or VC samples are drawn from the same population as the NV sample is fairly low (P $<$ 0.03), however there is a very strong probability (P = 0.847791) that the VS sample is drawn from the same population as the NV sample.  This suggests that there is a sharp distinction between the galaxy populations in the deepest underdensities and those in more moderately underdense regions.  This is most apparent in the large population of low luminosity galaxies found preferentially in the void center (see Section \ref{sec:excess}).  

\subsection{Luminosity Function}
\begin{figure}
\centering
\includegraphics[width=4in]{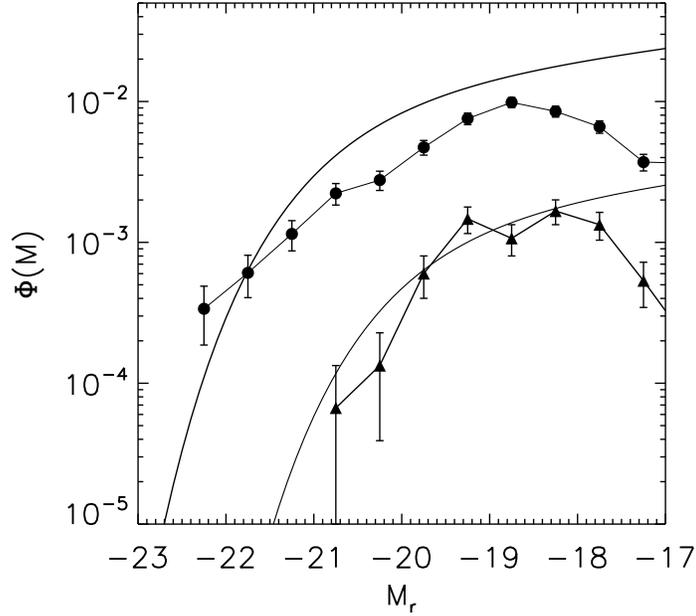}
\caption{Luminosity function of simulated void and `wall' galaxies.  Following \cite{Hoyle2005}, we split our sample into void galaxies with $\delta\rho/\rho < -0.4$ (triangles) and identify the remainder as `wall' galaxies (circles), with the respective observationally determined Schechter function overplotted.  The simulated void galaxies well reproduce observations, and we underpredict the function in the "wall" sample as we our simulation volume excludes higher density regions.   
 \label{fig:lumfun}}
\end{figure}

As the most luminous galaxies have been more robustly simulated throughout their merger history, we expect that this population will most closely match observations.  Indeed, the observed void galaxy luminosity function is very well reproduced (Figure \ref{fig:lumfun}).  Here we have divided our sample into two at $\delta\rho/\rho < -0.4$, following \cite{Hoyle2005}, and contrast the lower density void galaxies with the higher density `wall' galaxies.  Our simulated void sample very closely matches their observationally determined Schechter function fit.  Our `wall' sample contains too few galaxies, however our volume excludes any significant high density regions
and a significant number of  galaxies contaminated by coarse particles.  The turnover in both samples at $M_r = -18.5$ indicates the completeness limits in our simulated galaxy samples, and we expect that increased resolution would not significantly change our results for galaxies above this luminosity threshold. 

\subsection{Colors}
Compared to observations, our simulated void galaxy colors 
appear consistent, 
however they do not differ greatly from the non-void galaxies. Table \ref{tab:comp}  compares the mean properties of our galaxy samples with the photometric study by \cite{Rojas2004}, who construct a void galaxy sample of those galaxies with a third nearest neighbor distance greater than 7 $h^{-1}$ Mpc and a wall sample of the remaining galaxies.  
Duplicating the magnitude limit ($-17.77 > M_r > -22.5$) for their distant galaxy sample, and subsequently dividing it into a bright ($M_r < -19.5$) and faint ($M_r > -19.5$) subsamples, we see no consistent trend with density in our galaxy colors, but a strong dependence on magnitude.   
In all magnitude limited samples, we note that the NV galaxies are too blue when compared to the \cite{Rojas2004,Rojas2005} wall sample.

\begin{figure}
\centering
\includegraphics[angle=90,height=2.3in]{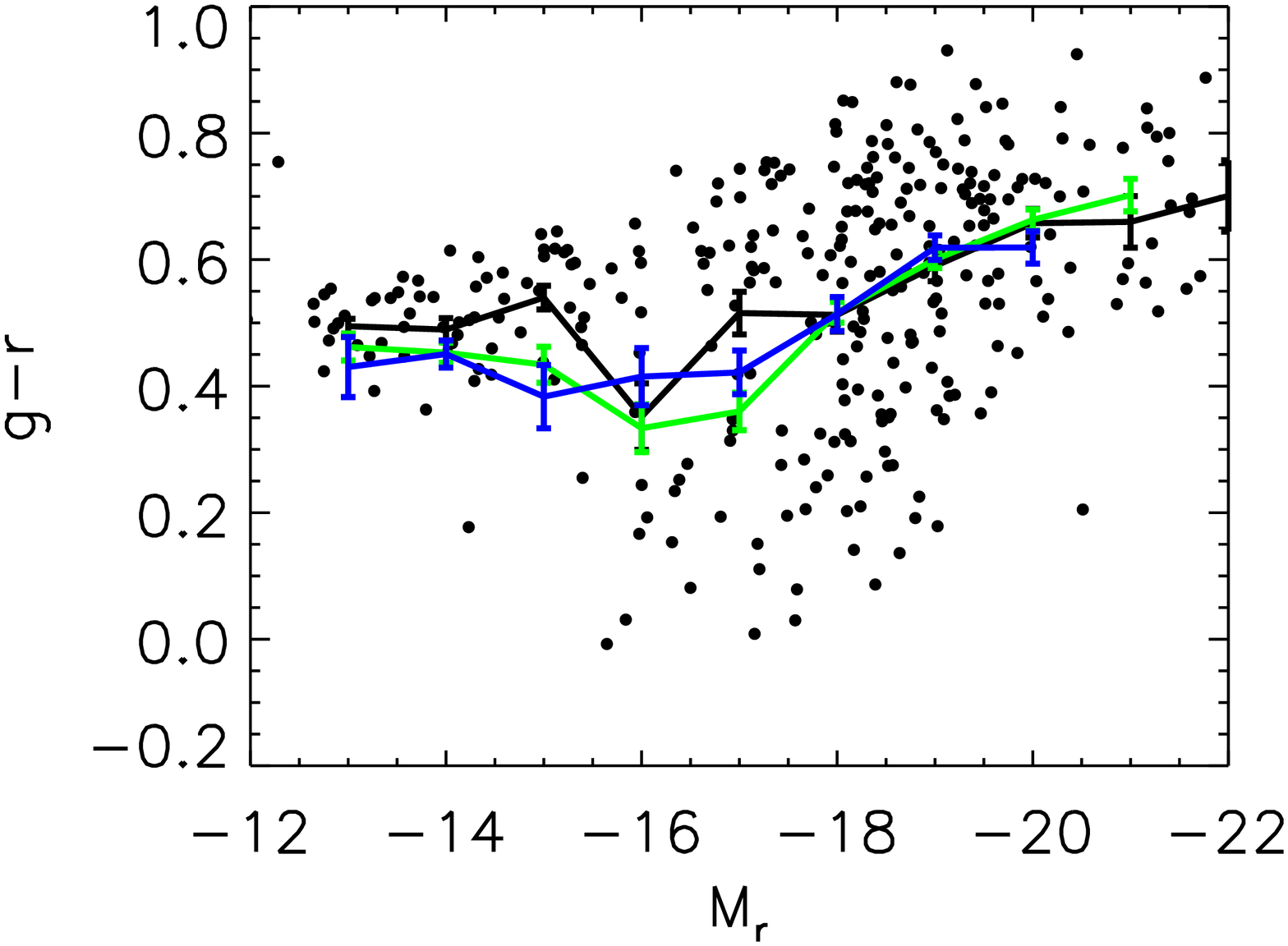}
\includegraphics[angle=90,height=2.3in]{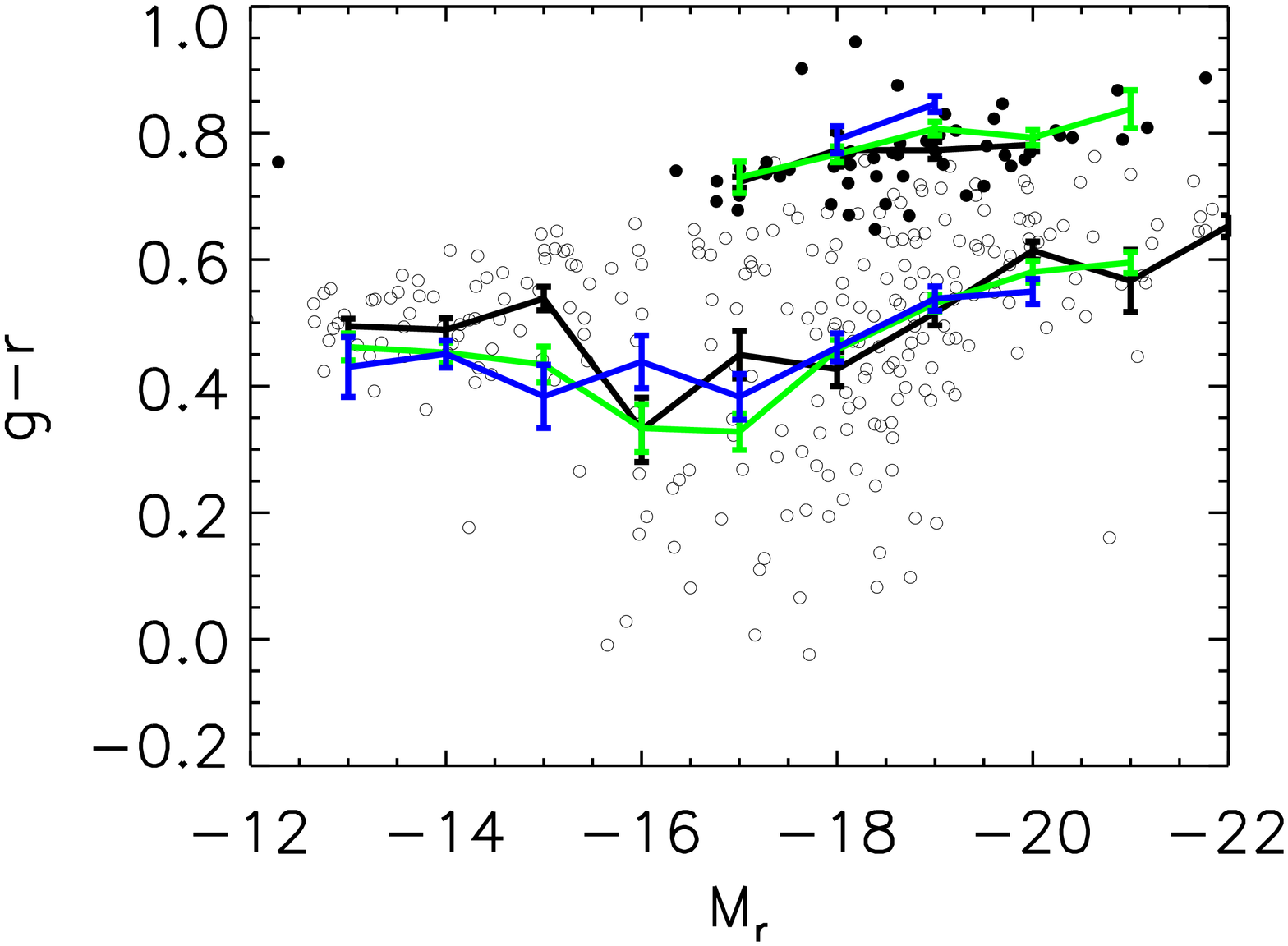}
\caption{Color magnitude diagram for the NV.  On the right, the sample has been split into red (filled circles) and blue (open circles) galaxies following the $u-r = 2.2$ cut identified by \cite{Strateva2001}.
In each, the mean for the NV galaxies is overplotted in black.  Also shown is the mean for galaxies in the void outskirts (0.5 $<$ delta $<$ 1.0, in green) and the mean for galaxies in the deepest underdensities (delta $<$ 0.5, blue). Among the dwarf galaxies, the void galaxies are somewhat bluer than the galaxies in average density environments. \label{fig:blueshift}}
\end{figure}

 In addition, none of the samples exhibit bimodality of galaxy colors with a red sequence and blue cloud (Figure \ref{fig:cmd}) such as is observed in both void and field galaxies \citep{BendaBeck2008}.  We compare the LVS with the geometrically selected Void Galaxy Survey (VGS, Kreckel et al. 2011, in preparation; selection criteria described in \cite{Kreckel2011}), and the NV with a magnitude-limited sample of SDSS galaxies selected at redshifts $0.1 < z < 0.3$.   Both comparisons show general agreement between simulation colors and observations down to the observational limit $M_r \sim -16$. 
 The choice of a lower scaling factor, $F$, in Equation  \ref{eq:avsim} corresponds to more extinction and redder galaxies, however it affects void and non-void colors similarly, does not affect the bimodality, and produces an excess of low-luminosity red galaxies compared to the VGS.  The lack of bimodality in color in the NV sample (Figure \ref{fig:blueshift}) suggests that the homogeneous distribution of galaxy colors is independent of density.  

While the absolute value calculated for our galaxy colors may not be entirely reliable, we may still consider relative differences between the void and non-void samples. 
\cite{BendaBeck2008} detect a slight blueward shift of the blue cloud galaxies that are deeper inside the void at fixed luminosity.   Considering the mean color at fixed luminosity (Figure \ref{fig:blueshift}, left) we see no such shift at the high-end luminosities they were considering ($-19.2 > {\rm M}_B > -20$).
They have fit their color distribution with two gaussians, which we cannot replicate as we see no bimodality, however dividing the sample at $u - r = 2.2$ \citep{Strateva2001}, we still find no trend with density (Figure \ref{fig:blueshift}, right).  We do see indications of a blue shift at fainter magnitudes, M$_r > -18$.  This is presumably due to the higher rates of star formation in galaxies at these densities, described below.

\subsection{H I Properties}
The H \textsc{i} distribution typically forms a disk, however the $\sim$5 kpc resolution 
used for the extraction of the hydro variables
is insufficient to attempt any analysis of the internal gas disk kinematics.  The H \textsc{i} surface density profiles we find rise too steeply at the outskirts of the disks, often achieving  fairly high column densities of $\sim10^{21} {\rm cm}^{-2}$ far outside the stellar disk (Figure \ref{fig:simobs}, right) (cf \citealt{Swaters2002}). 
This inaccurate density profile persists even if we are careful to include in the integrated H \textsc{i} map only that H \textsc{i} that is above a detectable level ($1 \times 10^{19}$ cm$^{-2}$) for modern instruments. We believe our measured H \textsc{i} masses to be an upper limit based on two limitations with this simulation.  First, the conversion of H \textsc{i} to H$_2$ through formation of H$_2$ on dust grains was not properly modeled even at this resolution.  This should result in an overestimate of the H \textsc{i} mass in the highest density, central regions of the gas disk.  Second, interstellar UV ionizing radiation by individual galaxies was neglected, which would have reduced H \textsc{i} in the outskirts of the disk, because it is likely that local ionizing radiation is larger than the meta-galactic background.

\begin{figure}
\centering
\includegraphics[angle=90,width=3in]{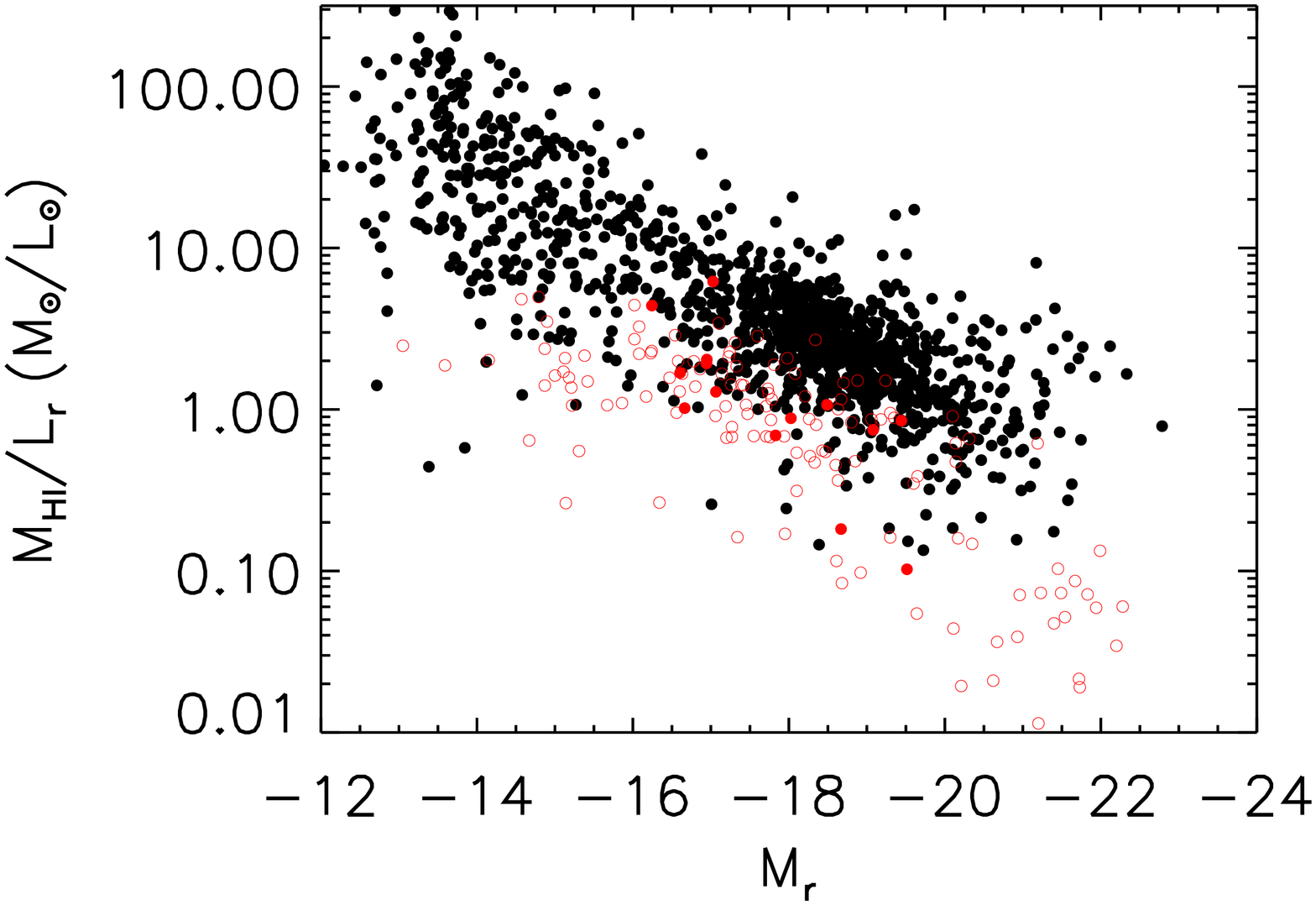}
\includegraphics[angle=90,width=3in]{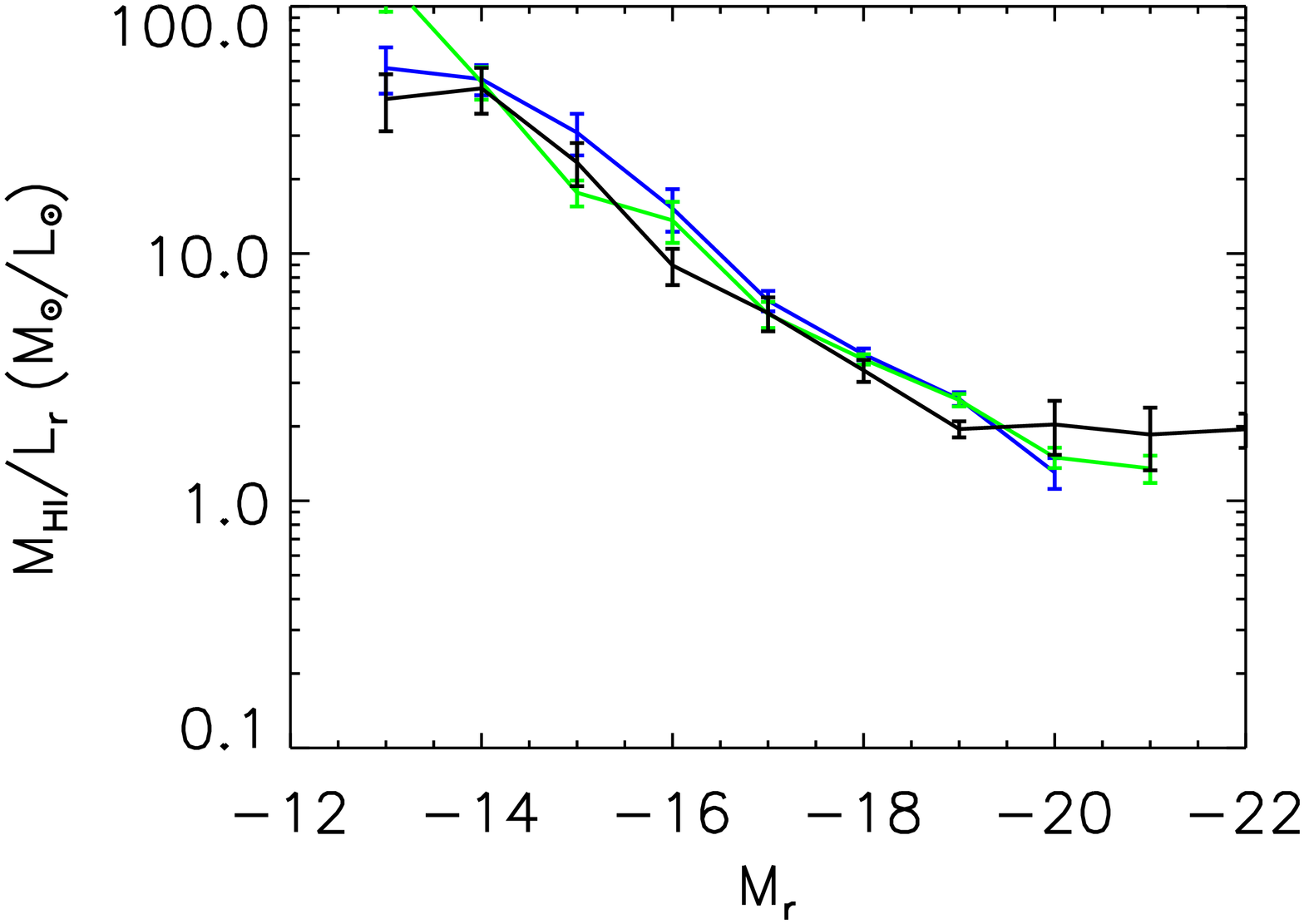}
\caption{H \textsc{i} mass to $r$-band light ratio.  On the left,  all simulated galaxies (black) are roughly an order of magnitude too high compared to the observed relation (in red) from \cite{Verheijen2001} and \cite{Swaters2002} (open circles) and \cite{Kreckel2011} (filled circles). On the right, the mean value for the LVS (blue),  NV (black) and VS (green) shows a very slight trend with density within the simulated dwarf galaxies. . \label{fig:mhilight}}
\end{figure}

\begin{figure}
\centering
\includegraphics[width=4in]{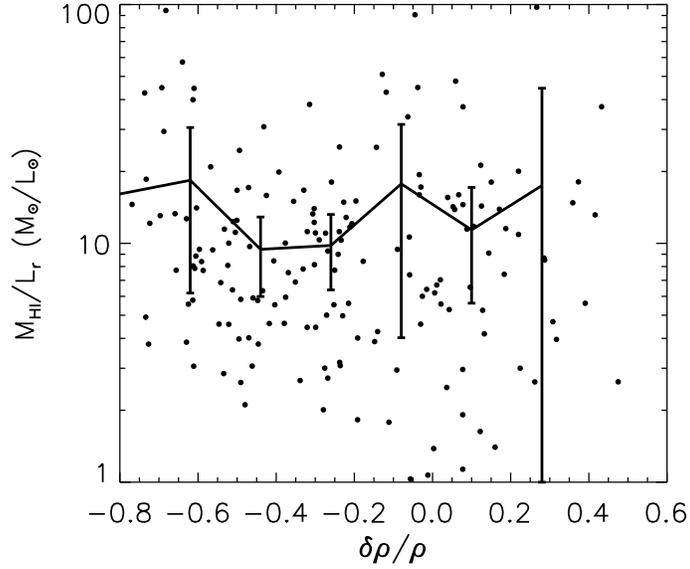}
\caption{M$_{\rm H \textsc{i}}$/L$_r$ vs density for all galaxies with $-15 > $M$_r > -17$, with the mean overplotted, reveals no clear trend with density. \label{fig:mhilightdens}}
\end{figure}

The resulting overestimate of the H \textsc{i} mass is apparent in comparing the H \textsc{i} mass to light ratio to observations (Figure \ref{fig:mhilight}, left).  It shows the same decreasing trend with increasing luminosity, however the overall value is about a factor of three too high compared with typical galaxies \citep{Verheijen2001, Swaters2002} or with void galaxies \citep{Kreckel2011}.    
We note that there is a slight trend with density among our three density-selected samples (Figure \ref{fig:mhilight}, right), however when we examine the specifics of the trend, particularly for low luminosity and dwarf galaxies (Figure \ref{fig:mhilightdens}), it is not very pronounced and certainly much weaker than the   trend observed for dwarf galaxies by \cite{Huchtmeier1997}. We see no trend as a function of distance from the void center for our VC galaxies.  

\subsection{Star Formation and Stellar Properties}
\begin{figure}
\centering
\includegraphics[angle=90,width=5in]{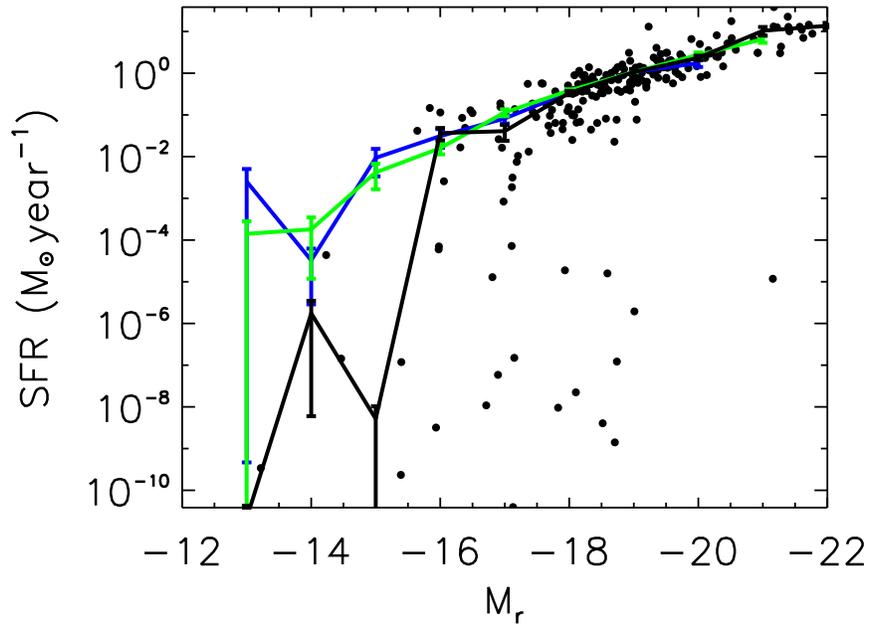}
\caption{SFR for NV galaxies (in black), with the mean overplotted for the LVS (blue),  NV (black) and VS (green). Only at the faintest luminosities is there any difference in the three samples. 
\label{fig:sfr1}} 
\end{figure}

\begin{figure}
\centering
\includegraphics[angle=90,width=3in]{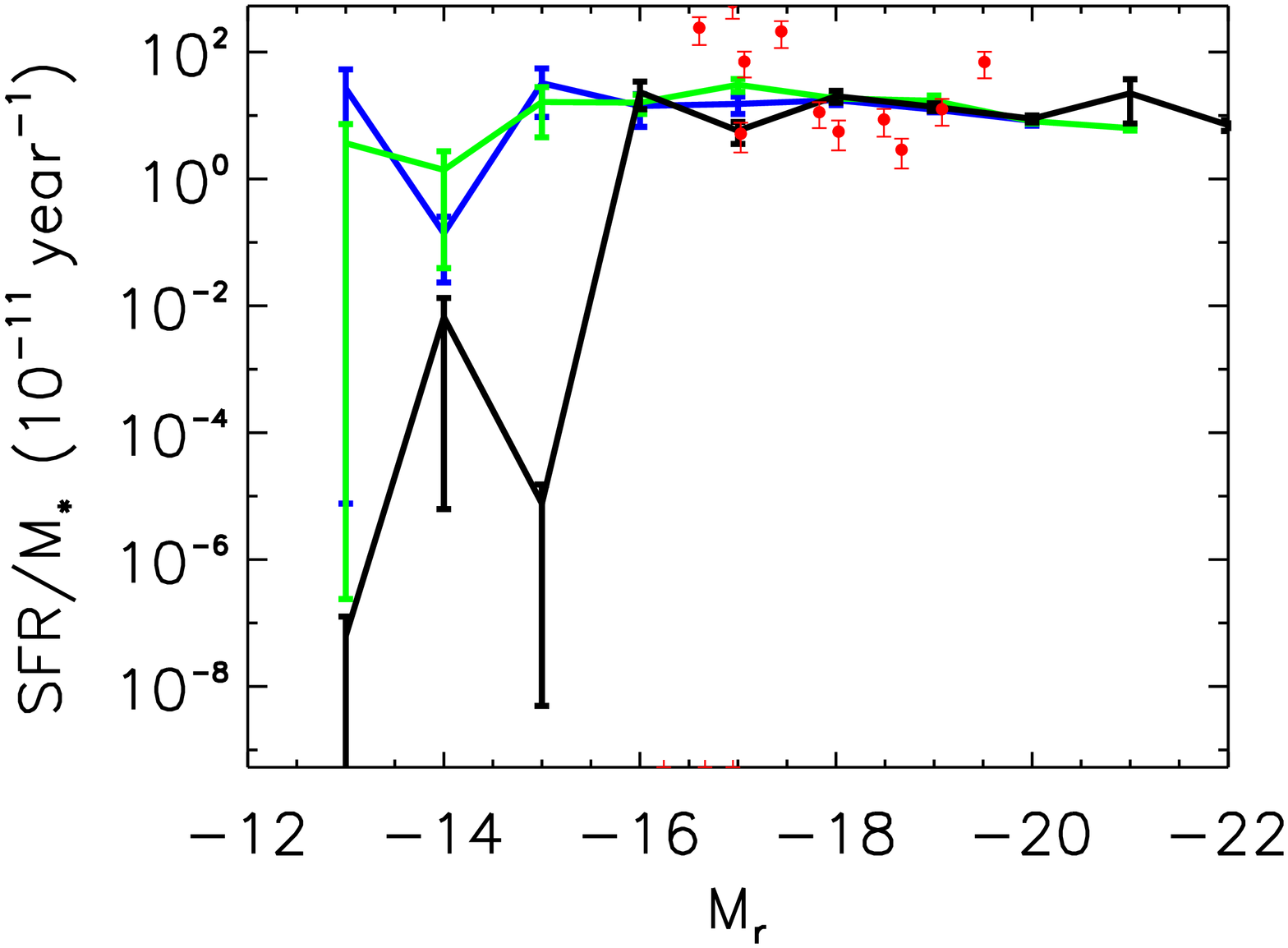}
\includegraphics[angle=90,width=3in]{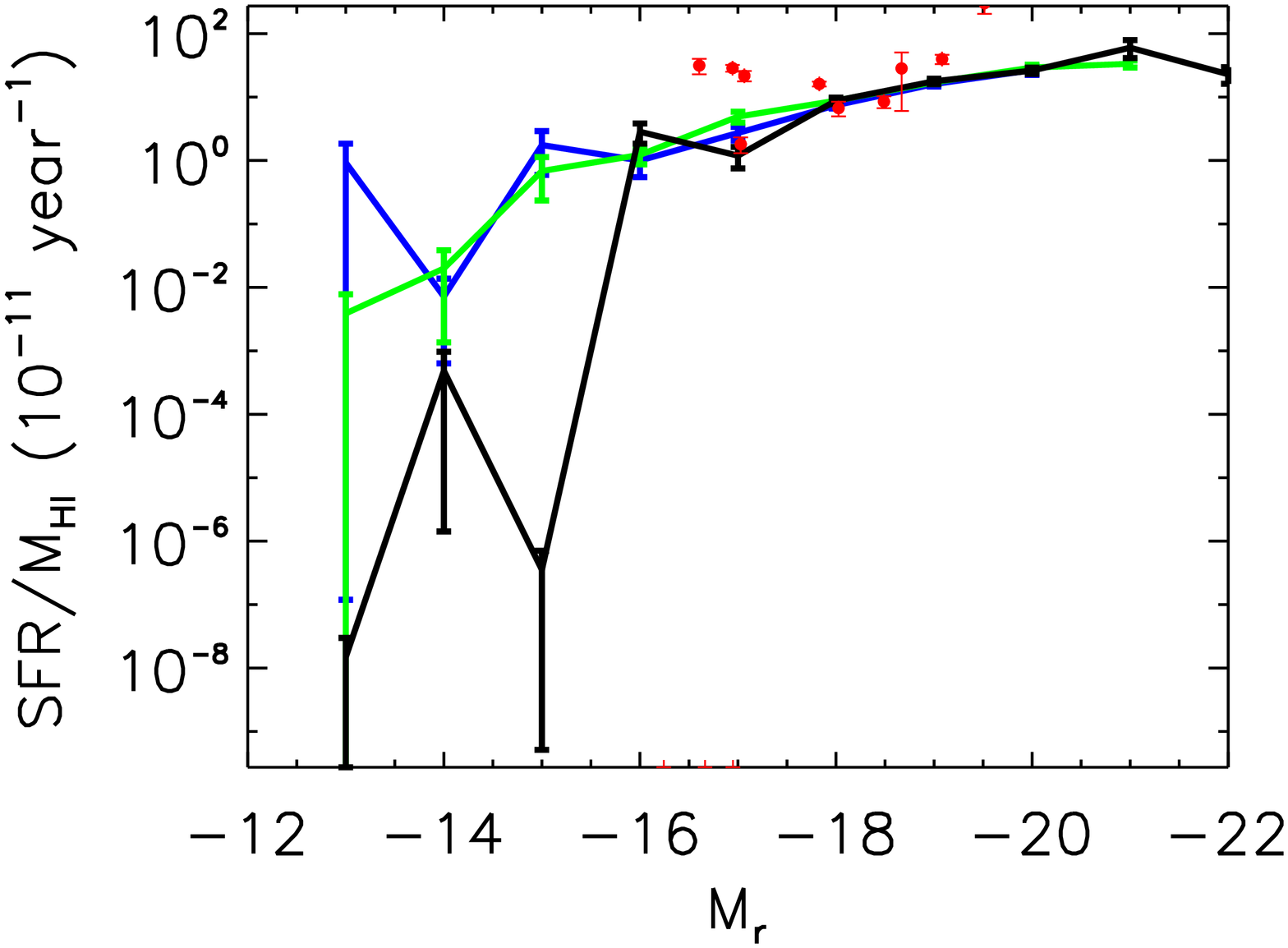}
\caption{S-SFR (left) and SFR/H \textsc{i} mass (right).   The LVS (blue) and VS (green) are somewhat higher than the NV (black) for the faintest galaxies, but consistent with each other and with the void galaxies observed by \cite{Kreckel2011} (red circles). The rather large discrepancy between observed and simulated values in the SFR per H \textsc{i} mass are consistent with the large discrepancy in the H \textsc{i} masses (see Figure \ref{fig:mhilight}).  \label{fig:sfr2}} 
\end{figure}

Although
there is a no strong change in the star formation rate (SFR) at lower densities (Table \ref{tab:comp}), it is clear from Figure \ref{fig:sfr1} that there is a strong dependence of SFR on luminosity. 
Thus we also consider the 
SFR scaled to the stellar mass (specific SFR, S-SFR), which is  significantly increased for observed void galaxies, but unchanged within the error in the mean for our simulation (Table \ref{tab:comp}).  We do, however, see significantly higher SFR and S-SFR for those lowest luminosity dwarf galaxies with M$_r > -16$ in underdense regions when compared to our NV galaxies (Figure \ref{fig:sfr1} and Figure \ref{fig:sfr2}, left).
The simulation does not reproduce the observed trend for increased SFR per H \textsc{i} mass at lower densities \citep{Kreckel2011}, and there is good agreement between densities for all but the faintest dwarf galaxies (Figure \ref{fig:sfr2}), however our H \textsc{i} masses are only upper limits and our sample is not complete for low mass galaxies.

\begin{figure}
\centering
\includegraphics[width=3.5in]{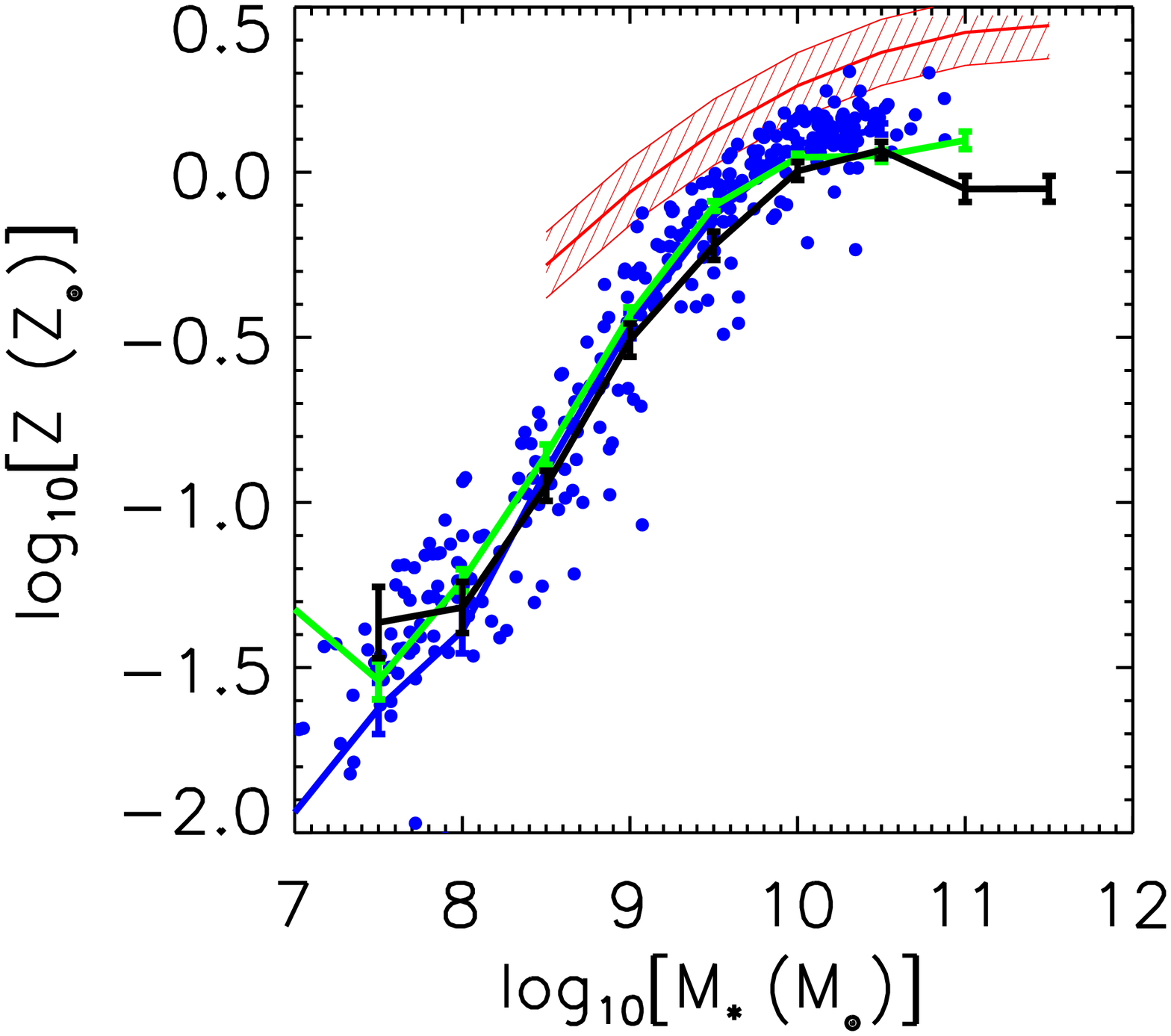}
\caption{Mass weighted metallicity as a function of stellar mass for the LVS (blue) with the mean overplotted for LVS,  NV (black) and VS (green).  We reproduce the same trend as \cite{Tremonti2004} (red) for increasing metallicity with increasing stellar mass, but find no trend with density.  The observed relation, normalized to Z$_\sun$ at 12 + log(O/H) = 8.69 \citep{AllendePrieto2001} with 1$\sigma$ error bars indicated by the shaded region, is somewhat higher as the measurements were done in the galaxy centers were metallicity is typically enhanced.  We reproduce the mass-metallicity relation for the Milky Way, with Z$_\sun$ at M$_* = 5 \times 10^{10} M_\sun$, and the LMC, with Z$_\sun / 2$ at M$_* = 5 \times 10^9 M_\sun$.
\label{fig:metals}}
\end{figure}

We find no trend in the mass-metallicity relation with density (Figure \ref{fig:metals}), or with distance from the void center.  This is likely because metal enrichment is mostly due to self-enrichment, and thus depends more on halo mass than large scale environment.
We reproduce the observed increase in metallicity for more massive galaxies \citep{Lequeux1979, Garnett1987, Zaritsky1994}.  We also reproduce the general shape in the mass-metallicity relation as calculated using the gas-phase oxygen abundance measurement of SDSS galaxies \citep{Tremonti2004}, however their survey measures only the 3$^{\prime\prime}$ of the observed galaxies, $\sim$25\% of the total light, in the central region where galaxy metallicity is typically higher \citep{Searle1971, Vilchez1988}.  This aperture bias may result in an overestimate of the total metal abundance by as much as a factor of 2, though it should not affect the general trend \citep{Tremonti2004}.  In comparing with their measured oxygen abundance, 12 + log(O/H), we have normalized the \cite{Tremonti2004} relation to solar metallicity, Z$_\sun$,  at 8.69 \citep{AllendePrieto2001}.  Our measured values reasonably well reproduce the Milky Way, with Z$_\sun$ at M$_* = 5 \times 10^{10} M_\sun$, and the LMC, with Z$_\sun / 2$ at M$_* = 5 \times 10^9 M_\sun$.   

\subsection{Excess of Low Luminosity Galaxies}
\label{sec:excess}
\begin{figure}
\centering
\includegraphics[height=2.1in]{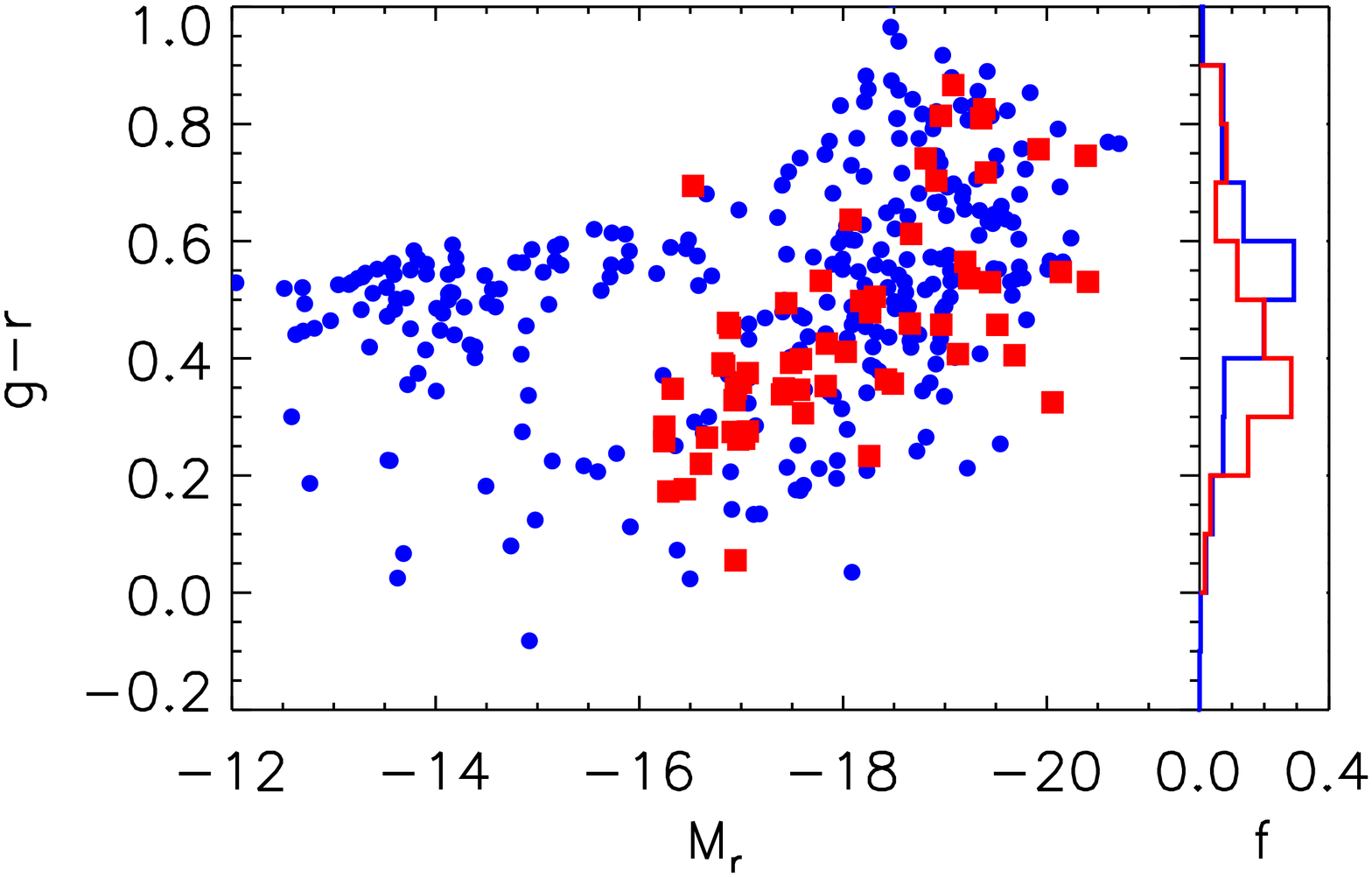}
\includegraphics[height=2.1in]{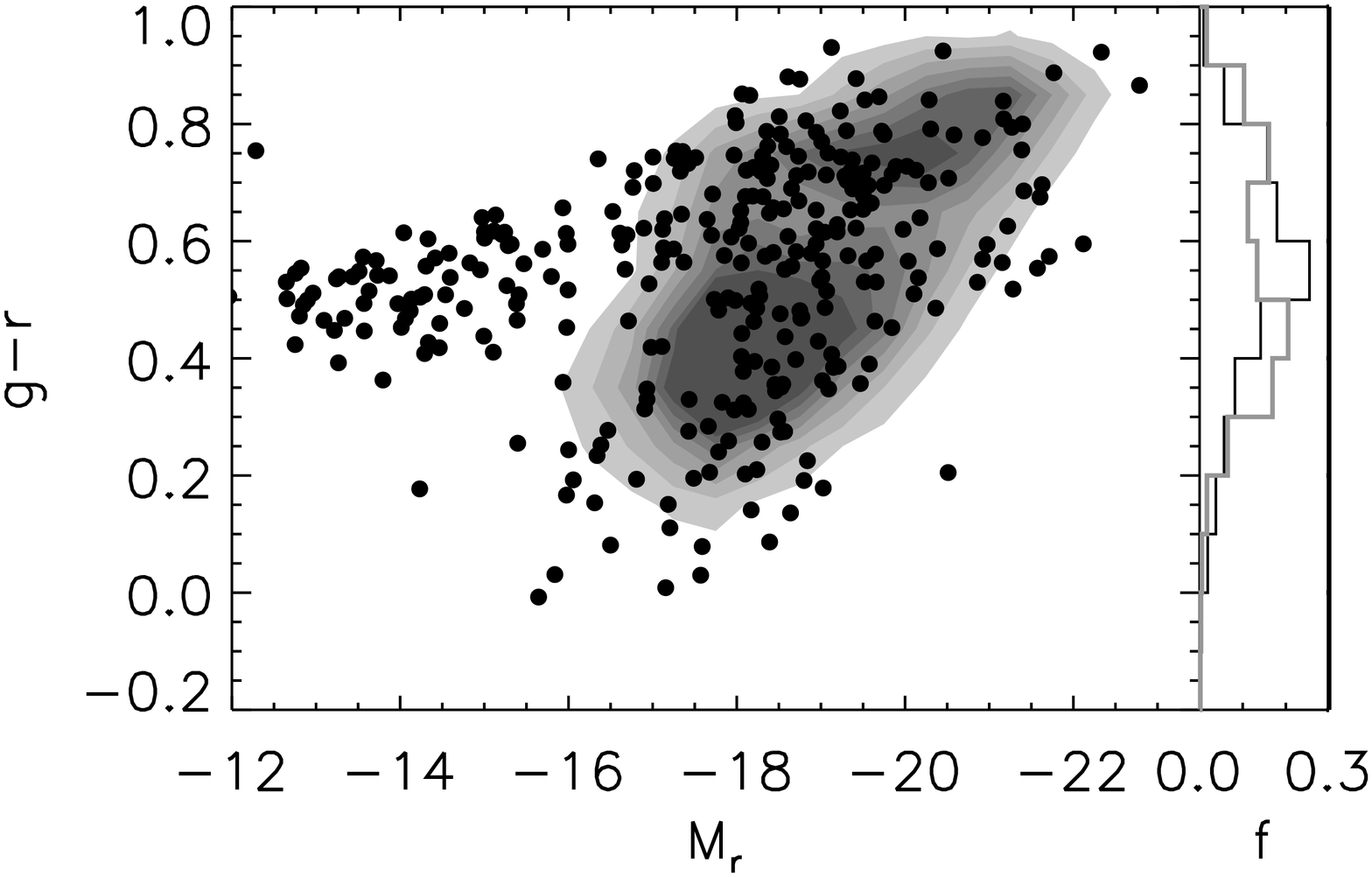}
\caption{Color magnitude diagram.  On the left, the LVS (blue circles) is in reasonable agreement with the observed VGS sample (red squares). On the right,   the NV (black circles) roughly traces the SDSS distribution for galaxies at redshifts $0.1 < z < 0.3$.   We see no bimodality in the colors of either simulated galaxy sample such as is observed in void and field galaxies \citep{BendaBeck2008}.  \label{fig:cmd}}
\end{figure}

\begin{figure}
\centering
\includegraphics[angle=90,width=4in]{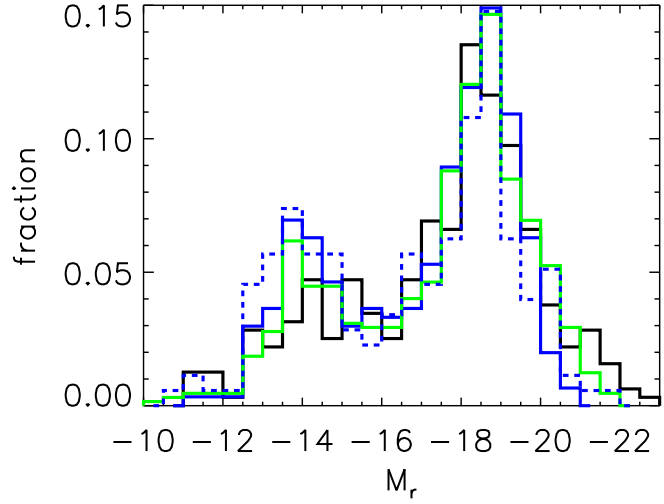}
\caption{Fraction of galaxies with a given luminosity for the VC (blue dashed line), LVS (blue),   NV (black) and VS (green). The excess of low luminosity galaxies is most apparent in the VC sample and relatively less dominant as a function of density , suggesting that this population is a result of the large scale underdensity. 
 \label{fig:histo}}  
\end{figure}

\begin{figure}
\centering
\includegraphics[angle=90,width=4in]{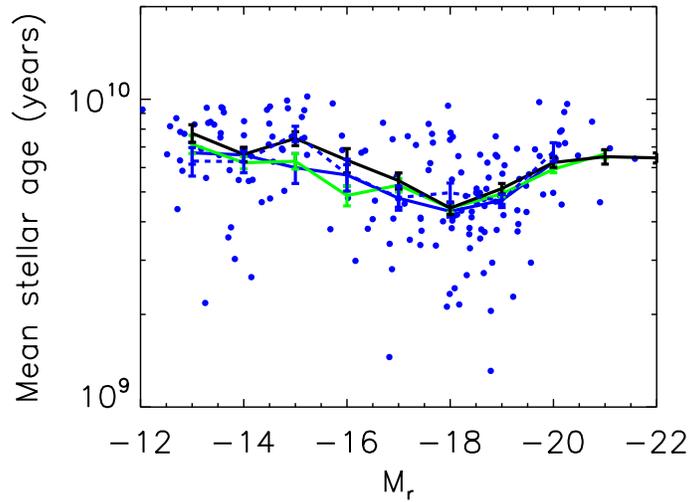}
\caption{Mean stellar ages as a function of luminosity for the LVS (blue) with the mean overplotted for LVS (blue solid), VC (blue dashed),  NV (black), and VS (green).  The low luminosity void galaxies are somewhat younger than the non-void galaxies.
 \label{fig:ages}}
\end{figure}

One surprising discovery  is the population of low luminosity galaxies apparent below the observational survey limit  in the color-magnitude diagram (Figure \ref{fig:cmd}). While these galaxies do inhabit the lowest mass dark matter halos, we believe that this excess is not a resolution effect as it persists into halos $\sim$3 times more massive than we can reliably resolve and has a strong trend with density.  Figure \ref{fig:histo} shows we have a slight excess of Mr $\sim -14$ galaxies for our NV, a noticeable excess for the VS, and a substantial excess for the LVS.  In the VC this excess is particularly notable.  This excess may be due in part to the youth of galaxies in the underdense regions (Figure \ref{fig:ages}), although at M$_r = -14$ the mean ages are about the same.  We note that while these galaxies are sufficiently resolved at $z=0$, the galaxies that formed them earlier in their merger history presumably were not, so their integrated properties are correspondingly less reliable than in more massive galaxies.  For example, these galaxies are generally gas rich, with masses $\sim10^9 M_\sun$, which may be due to decreased gas consumption at earlier times resulting from simulation resolution limits.

\begin{figure}
\centering
\includegraphics[width=4in]{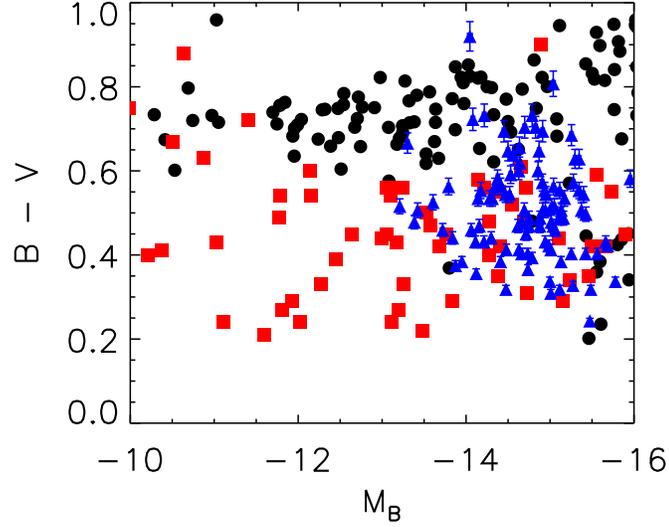}
\caption{Comparison of simulated NV low-luminosity galaxy colors with observational samples of dwarf galaxies from the Local Volume (red squares, \citealt{Makarova1999, Makarova2002, Makarova2009}) and from the SDSS (blue triangles, \citealt{Geha2006}).  Error bars reflect uncertanties in the color conversion from SDSS $ugzri$ bands to Johnson-Cousins UBVRI bands.  Simulated dwarf galaxies are distinctly redder than observed dwarf galaxies.
 \label{fig:dwarfs}}
\end{figure}

While complete volume limited samples of dwarf galaxies are not yet possible, comparison with observations show that our simulated dwarf galaxies are distinctly too red (Figure \ref{fig:dwarfs}).  Here we compare with dwarf galaxies within the Local Volume that have integrated $B$ and $V$ band observations \citep{Makarova1999, Makarova2002, Makarova2009}, and with dwarf galaxies identified within the SDSS redshift survey \citep{Geha2006}. We convert $ugzri$ colors for the SDSS and simulated galaxy samples to Johnson-Cousins $UBVRI$ colors following the prescription of \cite{Jester2005}.  In Figure \ref{fig:cmd} (right), when comparing our NV galaxy colors with typical SDSS galaxies, the simulated low luminosity galaxy population appears almost as an extension of the red sequence while the observed dwarf galaxies (Figure \ref{fig:dwarfs}) appear more continuous with the blue cloud.  While the observational samples are not complete in any way, there is no bias for color in their selection and both samples should be sensitive to a redder dwarf galaxy population such as we simulate.  This suggests that our simulated colors for the lowest mass galaxies, which contain very few star particles, are not realistic, presumably due to resolution effects earlier in their merger histories.

\begin{figure}
\centering
\includegraphics[width=2.2in,angle=90]{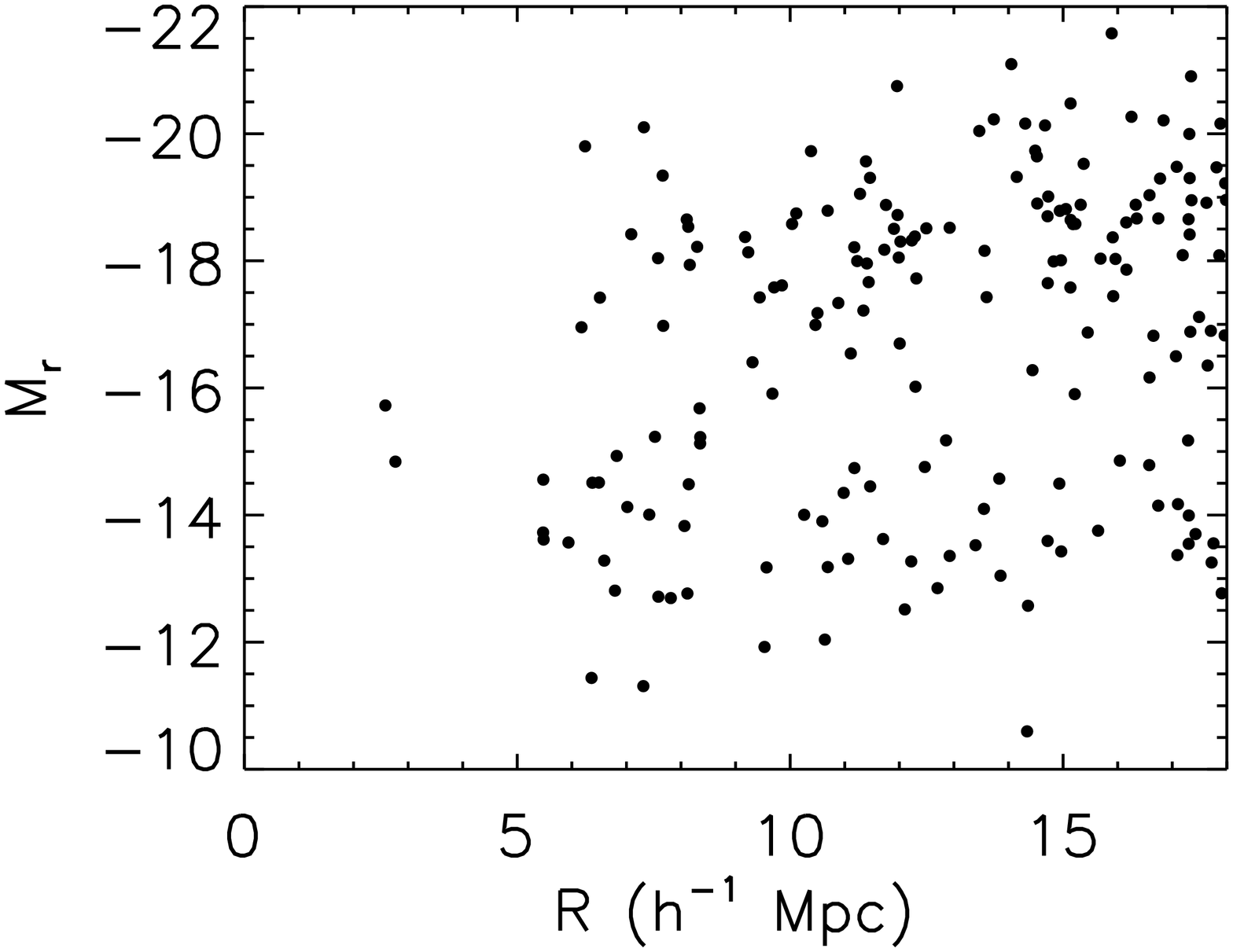}
\includegraphics[width=2.2in,angle=90]{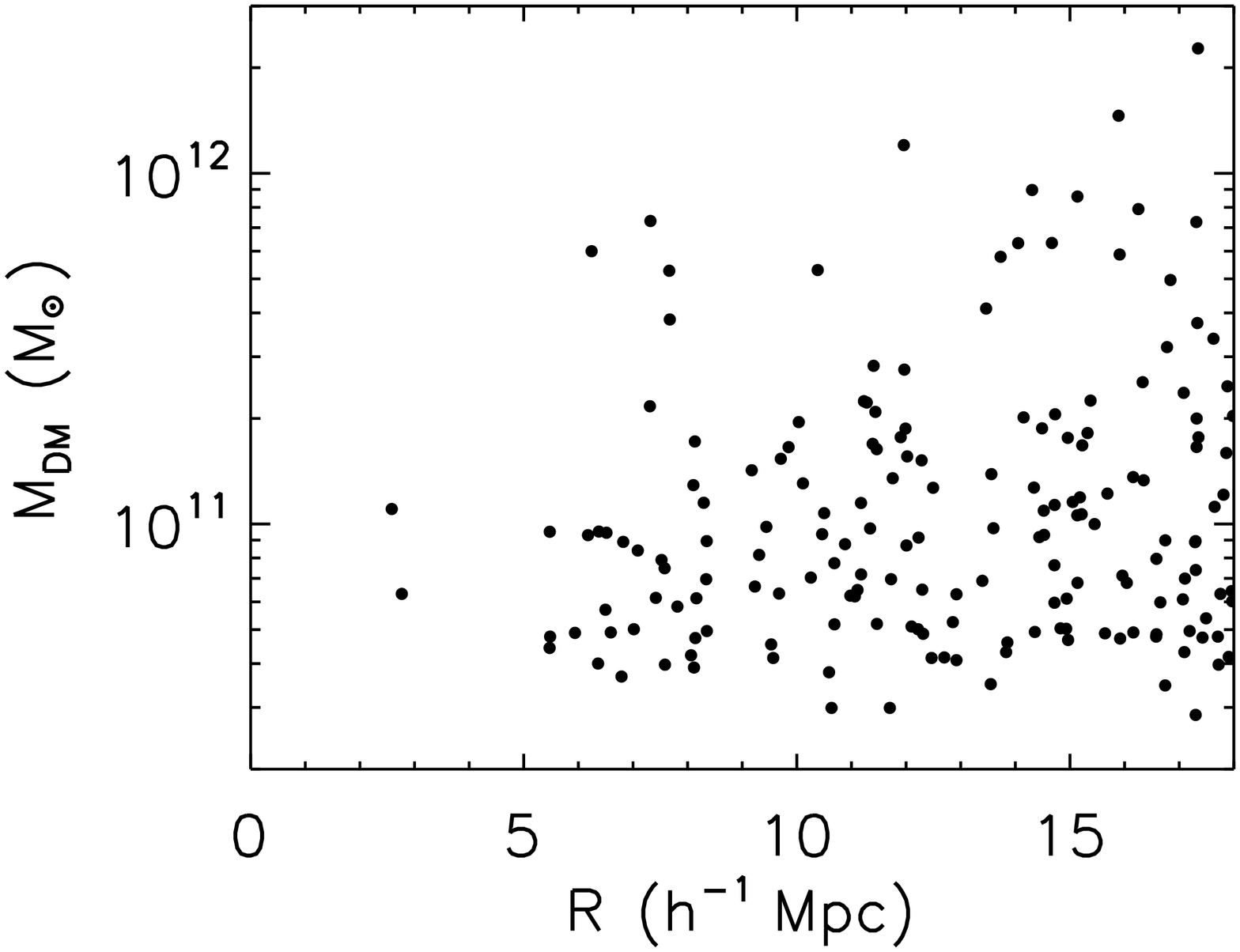}
\caption{M$_r$ (left) and M$_{DM}$ as a function of distance from the void center.  We see no trailing of low luminosity galaxies into the void, however the dark matter halos do seem to segregate by mass.
\label{fig:tinker}}
\end{figure}

Despite the discrepancy in colors, the increased number of simulated dwarf galaxies in the lowest density regions may present a way to reduce the discrepancy between the number of dark matter halos predicted within voids and the number of galaxies observed \citep{Peebles2001}. 
\cite{Tinker2009} suggested a sufficiently tailored halo occupation distribution could solve this problem, and predict that the maximum halo mass (and thus luminosity) will increase as a function of distance from the void center.  We see no such relation for VC galaxy luminosities, though we do see a slight trend with dark matter halo mass but not nearly as steep a cutoff (Figure \ref{fig:tinker}).

Comparing the number density of void galaxies in our simulation, which likely a lower limit due to imperfect resolution, with the most sensitive observational surveys we still over-predict the number of void galaxies.  If we consider observations of the Local Volume, where  the galaxy sample within 8 Mpc of us is thought to be fairly complete to M$_B$ = -12 \citep{Tikhonov2009}, and consider its overlap with the Local Void, which has a center 23 Mpc away from us and extends roughly 20 Mpc in radius \citep{Tully2008}, we find one void galaxy per 394 Mpc$^3$.  In our simulation the luminosity limit is lower, but from our VC sample we expect a similar volume should yield about 6 galaxies.  This is somewhat smaller than the factor of ten discrepancy found by \cite{Tikhonov2009}, but still quite pronounced.  It is still possible that these dwarf galaxies in the voids are low surface brightness and thus have been largely missed by existing surveys.  Future simulations with increased stellar resolution may be able to constrain the surface brightness of these systems.

The  one galaxy located in both the Local Volume and the Local Void, KK 246, has M$_B$ = -13.7, a color\footnote[1]{The mean color for our simulated low luminosity galaxies with M$_B > -16$ is $B-R \sim 1.3$.} of $B - R = 1$, and a significant H \textsc{i} mass (Kreckel et al. 2011, AJ submitted). Again it is too blue, but otherwise it is in good agreement with the expected properties of our low luminosity void galaxy population.  All sky H \textsc{i} surveys are on the verge of reaching the sensitivity needed to detect these galaxies.  HIPASS can detect galaxies with $10^8 M_\sun$ to distances of $\sim$10 Mpc, just within the edge of the Local Void.  ALFALFA reported no detections within the  Pisces- Perseus foreground void with a detection limit of $10^8 M_\sun$ \citep{Saintonge2008}, however considering the discrepancy between the H \textsc{i} masses of our simulated galaxies compared to observations, detailed examination of closer voids within 20 Mpc where we are sensitive to $\sim 10^7 M_\sun$ may be necessary.

\section{Conclusion}
\label{sec:conclusion}
We have run cosmological simulations using the adaptive mesh refinement
Eularian hydro code, Enzo, with a moderate resolution refined region $31 \times 31 \times 35 h^{-3}$Mpc centered on a large $\sim$ 30 $h^{-1}$ Mpc diameter void.  We have constructed mock observations of the integrated properties (color, star formation rate, gas content, etc.) in a sample of roughly 900 void galaxies, 300 in especially low density regions and 200 located well within the void center, and 
300 additional galaxies in the surrounding average density filaments.

We fairly accurately reproduce the range of galaxy luminosities and colors observed, but are unable to reproduce the observed bimodality in galaxy colors.  
At low luminosities (M$_r > -16$), where we more affected by resolution effect from earlier in the galaxy merger history and beyond the scope of most observational studies, we see that void galaxies are bluer with higher SFR and S-SFR, but we see no trend for the brightest galaxies (M$_r < -18$).   

The gas content of all our simulated galaxies is anomalously high, roughly 3 times what is expected for their luminosity when compared with observations, and their radial extent too large, both of which we 
interpret as a result of limitations in the implementation of chemistry and radiative transfer in the simulations. 
Taking that into account, we do reproduce the trend for fainter galaxies to have relatively more H \textsc{i} for their luminosity, but find no trend with density or distance from the void center.  We reproduce the general mass-metallicity trend observed in SDSS galaxies, and  we see no trend with density.

We do not see the strong segregation  of the most luminous galaxies to the void edges predicted by \cite{Tinker2009}, however we do see that the most massive dark matter halos avoid the void center.  We observe a significant population of low luminosity (M$_r \sim -14$) dwarf galaxies that is preferentially located in low density regions and specifically in the void center.   This population is too faint to be detected in most large scale redshift surveys, however studies of the Local Void may soon become sensitive to this population, where we do observe a relatively lower median luminosity  \citep{Nasonova2010}.  

\acknowledgments
KK thanks Jacqueline van Gorkom for useful discussions and valuable comments. This work was supported in part by the National Science Foundation under grant \#1009476 to Columbia University. This work is also supported in part by grants NNX08AH31G and NAS8-03060.
Funding for the SDSS has been provided by the Alfred P. Sloan Foundation, the Participating Institutions, the National Science Foundation, the U.S. Department of Energy, the National Aeronautics and Space Administration, the Japanese Monbukagakusho, the Max Planck Society, and the Higher Education Funding Council for England. The SDSS Web Site is http://www.sdss.org/.

\clearpage

\begin{table}
\begin{center}
\tabletypesize{\scriptsize}
\caption{Categories defined for comparison between simulated galaxy populations.
\label{tab:cats}}
\begin{tabular}{l l c c}
\tableline
\tableline
Category & Description &  Definition & Number of Galaxies \\
\hline
NV & non-void  & $0 < \delta\rho/\rho < 1$ & 318 \\
VS & moderate underdensity & $-0.5 < \delta\rho/\rho < 0$ & 648 \\
LVS & lowest density  & $\delta\rho/\rho < -0.5$ & 302 \\
VC & void center & R $< 18 h^{-1}$ Mpc & 176 \\
\tableline
\end{tabular}
\end{center}
\end{table}

\begin{deluxetable}{l c c c c c }
\tablewidth{0pt}
\tabletypesize{\scriptsize} 
\tablecaption{Average color and star formation properties of our simulated galaxies as compared with the magnitude limited samples of  \cite{Rojas2004} and \cite{Rojas2005}. \label{tab:comp}}
\tablehead{
\colhead{Property} & \colhead{LVS} & \colhead{VS} & \colhead{NV} & \colhead{Rojas Void} & \colhead{Rojas wall} 
}
\startdata
\multicolumn{6}{c}{ $-17.77 > $M$_r > -22.5$ }\\
 \hline
g-r & 	0.586 $\pm$ 0.0144 & 	0.600 $\pm$ 0.00930  &  0.589 $\pm$ 0.0133 & 0.615 $\pm$ 0.007 & 0.719 $\pm$ 0.002 \\
SFR & 	        0.988   $\pm$  0.0838 &      1.80  $\pm$    0.172 &      2.57   $\pm$   0.394 	  & 0.734 $\pm$ 0.025 & 0.747 $\pm$ 0.007 \\
S-SFR ($\times 10^{-11}$) & 
		12.9   $\pm$    1.17 &       15.0   $\pm$    1.65 &      14.3   $\pm$    1.80  & 3.744 $\pm$ 0.108 & 2.629 $\pm$ 0.034\\

\hline
\multicolumn{6}{c}{ $-19.5 > $M$_r > -22.5$ }\\
\hline
g-r &  	0.593  $\pm$ 0.0234 & 0.656 $\pm$   0.0145 &		0.666 $\pm$ 0.0169 &	0.686 $\pm$ 0.009 &  0.765 $\pm$ 0.002 \\ 
SFR &	      1.99    $\pm$  0.318 &      4.30   $\pm$   0.5309 &      6.54   $\pm$    1.01		& 1.136 $\pm$  0.063 &  0.920 $\pm$ 0.016 \\ 
S-SFR ($\times 10^{-11}$) & 
 		       8.31    $\pm$   1.02 &      8.30   $\pm$   0.753 &      12.1   $\pm$    4.09  & 3.133 $\pm$ 0.169 & 2.137 $\pm$ 0.004 \\

\hline
\multicolumn{6}{c}{ $-17.77 > $M$_r > -19.5$} \\
\hline
g-r &  	0.585  $\pm$ 0.0169 & 0.579 $\pm$   0.0113 &		0.550 $\pm$ 0.0171 &	0.567 $\pm$ 0.009 &  0.645 $\pm$ 0.003 \\ 
SFR &	0.770  $\pm$   0.0606 &     0.838  $\pm$   0.0544 &     0.618   $\pm$  0.0548	& 0.508 $\pm$  0.0237 &  0.530 $\pm$ 0.009 \\ 
S-SFR ($\times 10^{-11}$) & 
 		 13.9    $\pm$   1.39 &      17.6   $\pm$    2.25 &      15.3   $\pm$    1.79 & 4.146 $\pm$ 0.137 & 3.349 $\pm$ 0.005 \\
\enddata
\end{deluxetable}

\end{document}